\documentclass[aps,pra,twocolumn,superscriptaddress,showpacs,showkeys,amsmath,amssymb]{revtex4}

\usepackage{amsfonts}
\usepackage{amssymb,amsmath}
\usepackage{mathrsfs}
\usepackage{latexsym}
\usepackage{amsmath}
\usepackage[cp1251]{inputenc}
\usepackage{graphicx}
\usepackage{dcolumn}
\usepackage{bm}
\usepackage{color}

\RequirePackage{ifthen}
\RequirePackage[pdfstartview=FitH]{hyperref}
\begin{document}
	
	\title{Second root of dilute Bose-Fermi mixtures}
    \author{O.~Hryhorchak}
	\author{V.~Pastukhov\footnote{e-mail: volodyapastukhov@gmail.com}}
	\affiliation{Professor Ivan Vakarchuk Department for Theoretical Physics, Ivan Franko National University of Lviv, 12 Drahomanov Street, Lviv, Ukraine}

	\date{\today}

	
	
	\begin{abstract}
	We discuss an equilibrium mean-field properties of mixtures consisting of bosons and spin-polarized fermionic atoms with a point-like interaction in an arbitrary dimension $2<d<4$. Particularly, we discuss except the standard weak-coupling limit of the system with slightly depleted Bose condensate and almost ideal Fermi gas, the (meta)stable phase with dimers composed exactly of one boson and one fermion. The peculiarities of the fermion-dimer and the boson-dimer three-body effective interactions and their impact on the thermodynamic stability of the dilute Bose-Fermi mixtures are elucidated.
	\end{abstract}
	
	\maketitle
\section{Introduction}
Mixtures composed of mutually interacting Bose and Fermi particles attract attention of researches from the early days of the quantum liquids history. The phase diagram of the paradigmatic example -- the $^4$He-$^3$He mixture -- was extensively studied both theoretically \cite{Bardeen1967,Krotscheck1993} and experimentally \cite{Eselson1973,Bruyn_Ouboter1986}. Developments of last few decades in a field of cooling atoms of different species of alkalis \cite{Truscott2001,Schreck2001}, together with the possibility to control the parameters of their two-body potentials by means of the Feshbach resonances \cite{Park2012} have lead to the creation of the gaseous Bose-Fermi mixtures with an arbitrary strong interparticle interaction \cite{Stan2004,Inouye2004}. Although a simple perturbative analysis of the thermodynamic stability of these systems at a weak coupling was carried out long ago by Saam \cite{Saam1969}, and in more recent times of realization of ultracold quantum gases in Refs.~\cite{Viverit2000,Albus2002,Sogo2002,Viverit2002,Capuzzi2003}, the interest to the topic still does not go down \cite{Lous2018,Rakshit2019,Manabe2019,Zheng2021,Fritsche2021,Milczewski2021,Duda2021} in the community.

The present article deals with the dilute Bose-Fermi mixture with the tightly-bound dimers formed in dimensions between $d=2$ and $d=4$ at absolute zero. The previous studies in this context were mostly focused on the three-dimensional systems with the boson-fermion interaction fine-tuned to either narrow \cite{Avdeenkov2006} or broad \cite{Watanabe2008,Son2011,Kharga2017} Feshbach resonances. Nonetheless, there are some quantitative differences in physics of these two models, the general features of their properties are quite similar. Particularly, depending on the interaction strength, the mean-field phase diagram \cite{Marchetti2008} of the composite fermions at finite temperatures includes states with and without the Bose-Einstein condensate. For the imbalanced compositions of bosons and fermions, there are two Fermi surfaces with the manifested \cite{Powell2005} two Luttinger theorems in a phase with an unbroken global symmetry. These results mostly remain true when the Gaussian fluctuations are taken into account, and the ground state of the system possesses a rich phase diagram with a number of quantum phase transitions \cite{Ludwig2011} of various order. 

Likewise the Fermi-polaron problem \cite{Massignan2014}, the $t$-matrix approximation was shown to be quite accurate \cite{Guidini2015} for the description of the paired Bose-Fermi mixture even when the concentration of bosons is not small. This was explicitly verified in \cite{Bertaina2013} by comparison to the results of Monte Carlo simulations, and in Refs.~\cite{Fratini2010,Fratini2012} the interaction-induced breakdown of the Bose-Einstein condensation in the whole temperature range was investigated within this approach. In condensate phase, the composite-fermion propagator possesses two poles \cite{Sogo2013} of a different physical nature. Even richer structure of the quasiparticle excitations \cite{Fratini2013} is displayed by the single-particle fermionic and bosonic spectral functions in the normal phase, when the Bose condensate is totally depleted by the strong boson-fermion attraction. The common opinion is that the thermodynamic stability of the system requires some finite inter-bosonic repulsion \cite{Yu2011}, especially when the number of bosons exceeds number of fermions. 

An important aspect \cite{Cui2014} of the Bose-Fermi mixtures properties, less discussed in the literature, is the atom-dimer scatterings. From the point of view of the stability of a mixed state in a dilute system, the effective fermion-dimer repulsion stabilizes mixture with majority of fermions and boson-dimer attraction forces the collapse. Another feature of the three-body physics is the emergence of the universal Efimov effect \cite{Efimov} (see \cite{Naidon} for recent discussion). The latter is crucial for the thermodynamic properties of the Bose-Fermi mixture by making a phase with the boson-fermion dimers formed at least metastable. Here we attempted to address all these questions in order to explore in detail the stability regions of dilute systems.

\section{Contact two-body potential in arbitrary $d<4$}
Let us briefly overview the main ideas behind the contact ($\delta$-like) interaction between particles. This concept is quite useful when the range of a two-body potential is small and one is interested in the universal low-energy physics of the system. Then the contact interaction can be presented in a twofold way: the first one stipulates the introduction of a large ultraviolet (UV) cutoff $\Lambda$ (which is a reminiscent of a range of the realistic two-body potential), while the second one explicitly appeals to the (typically non-hermitian) $\delta$-pseudo-potential. Below, while considering the many-body behavior of the system, the former way will be of great practical use, therefore let us first consider a fermion of mass $m_f$ and boson of mass $m_b$ in a very large volume $L^d$ in $d$ dimensions with the periodic boundary conditions imposed. The appropriate Schroedinger equation for the relative motion of two particles reads
\begin{eqnarray}\label{Eq_2B}
-\frac{\hbar^2}{2m_r}\nabla^2\psi({\bf r})+g_{\Lambda}\delta_{\Lambda}({\bf r})\psi({\bf r})=\mathcal{E}\psi({\bf r}),
\end{eqnarray}
where $m_r=\frac{m_fm_b}{m_f+m_b}$ and $g_{\Lambda}$ are the reduced mass and the bare coupling, respectively, and the UV-cutoff-dependent delta-function should be understood as the following integral $\delta_{\Lambda}({\bf r})=\frac{1}{L^d}\sum_{|{\bf p}|<\Lambda} e^{i{\bf p}{\bf r}}$ in the momentum space. Being interested in the spherically symmetric ($s$-wave) bound states with energy $\mathcal{E}=\epsilon_B=-\frac{\hbar^2}{2m_ra^2}$, we can write down the non-normalized wave function
\begin{eqnarray}\label{psi_2B}
\psi(r)=\frac{1}{L^d}\sum_{|{\bf p}|<\Lambda}\frac{e^{i{\bf p}{\bf r}}}{p^2+a^{-2}}.
\end{eqnarray}
The requirement of comparability with the Eq.~(\ref{Eq_2B}) yields the condition on $g_{\Lambda}$
\begin{eqnarray}\label{g_Lambda}
g^{-1}_{\Lambda}+\frac{1}{L^d}\sum_{|{\bf p}|<\Lambda}\frac{2m_r/\hbar^2}{p^2+a^{-2}}=0.
\end{eqnarray}
An introduction of the renormalized (`observable') coupling constant
\begin{eqnarray}\label{g_def}
g^{-1}=g^{-1}_{\Lambda}+\frac{1}{L^d}\sum_{|{\bf p}|<\Lambda}\frac{2m_r}{\hbar^2p^2},
\end{eqnarray}
enables to eliminate the explicit dependence of $g_{\Lambda}$ on the UV cutoff in dimensions below $d<4$. In a higher $d$s the wave function (\ref{psi_2B}) is not square normalizable. By plugging (\ref{g_def}) in (\ref{g_Lambda}) and computing the sum over wave vector in the limit of free space ($L\to \infty$), one relates a positive $g$s to the width of a binding energy of two atoms in vacuum
\begin{eqnarray}\label{g}
g^{-1}=-\frac{\Gamma(1-d/2)}{(2\pi)^{d/2}}\left(\frac{m_r}{\hbar^2}\right)^{d/2}|\epsilon_B|^{d/2-1}.
\end{eqnarray}
The simplicity of potential energy in Eq.~(\ref{Eq_2B}) also allows the exact calculations of the scattering states. Particularly, the computed to all orders of perturbation theory energy shift $\mathcal{E}=g/L^d+\mathcal{O}(1/L^{2d})$ of the lowest level (${\bf p}={\bf 0}$) reveals the nature of the contact two-body interaction: depending on a sign of the renormalized coupling, the potential behaves either like the repulsive ($g>0$) or the attractive ($g<0$) one (recall that a single $s$-wave bound state appears for positive $g$s, which may be somewhat misleading). 
Another important lesson to be learned from the above analysis is that the renormalized coupling $g$ should be identified as the physical one, i.e. determines the strength of the two-body interaction and all observables.

\section{Standard weak-coupling consideration}\label{weak-coupling_sec}
\begin{figure}[h!]
	\centerline{\includegraphics
		[width=0.45
		\textwidth,clip,angle=-0]{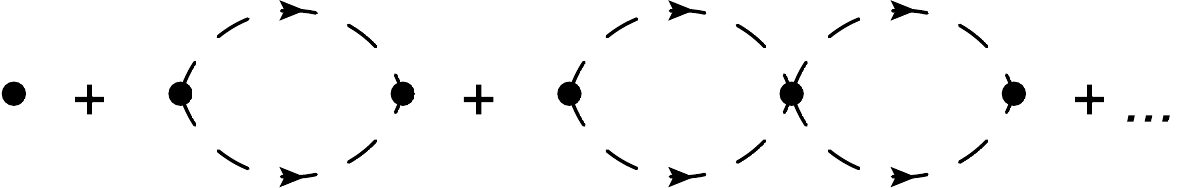}}
	\caption{The simplest diagram with four condensate lines (not shown) contributing to $\Omega_{MF}$. Dashed lines denote the bare bosonic propagators $G_b(K)=\frac{1}{i\omega_k-\xi_b(k)}$ written down here in a $d+1$ momentum space [from now on $K=(\omega_k,{\bf k})$], and dots stand for the bosonic zero-order vertices $g_{b,\Lambda}$. Strictly speaking, loops are infrared convergent only for $\mu_b\le 0$, but apriory knowing that $\mu_b$ is small we neglect it here. The latter is only allowed in $d>2$. The series sums up in the geometric progression, where the UV dependence of loop exactly cancels the $\Lambda$-dependent term in $g_{b,\Lambda}$ (\ref{g_def}) out, giving a finite result announced in (\ref{Omega_MF}).}\label{g_b_fig}
\end{figure}

\begin{figure}[h!]
	\centerline{\includegraphics
		[width=0.45
		\textwidth,clip,angle=-0]{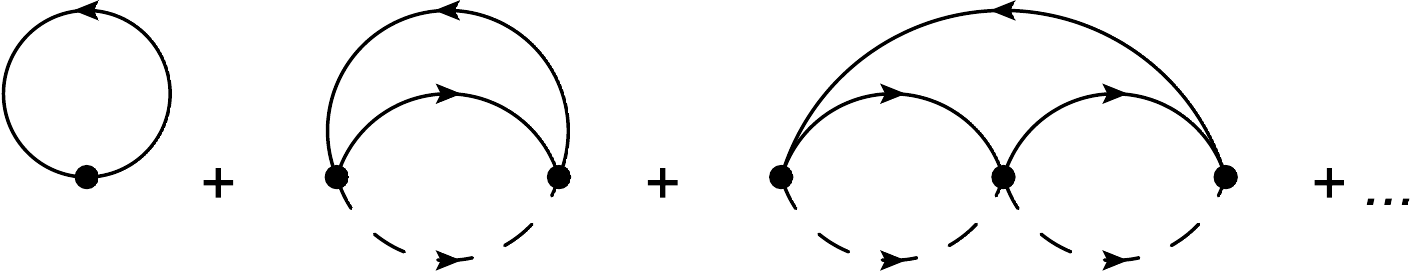}}
	\caption{Series determining the last term in $\Omega_{MF}$. Here solid lines represent the fermionic propagators $G_f(P)=\frac{1}{i\nu_p-\xi_f(p)}$ and dots stand for $g_{\Lambda}$. Note that in (\ref{Omega_MF}) only the leading-order term at small $g$s is kept.}\label{g_f_fig}
\end{figure}

Although below only limit of zero temperature will be considered, for the discussion of the thermodynamic limit of the Bose-Fermi mixture we adopt the Euclidean-time path integral approach with an action
\begin{eqnarray}\label{S}
S=\int dx\,b^*\left\{\partial_{\tau}-\xi_b\right\}b-\frac{g_{b,\Lambda}}{2}\int dx\,|b|^4\nonumber\\
+\int dx\,f^*\left\{\partial_{\tau}-\xi_f\right\}f-g_{\Lambda}\int dx\,f^*f|b|^2,
\end{eqnarray}
where $x=(\tau, {\bf r})$ is the short-hand notation for position ${\bf r}\in L^d$ and an imaginary time $\tau \in [0, \beta)$ (with the inverse temperature $\beta\to \infty$ in the final formulas); $\xi_{b,f}=\varepsilon_{b,f}-\mu_{b,f}=-\frac{\hbar^2\nabla^2}{2m_{b,f}}-\mu_{b,f}$ and $\mu_{b,f}$ are the chemical potentials that fix the average densities $n_{b,f}$ of bosons and fermions in the system. The boson-boson bare coupling $g_{b, \Lambda}$ can be related to its `observable' counterpart $g_b$ in a way described in the previous section. It is also understood that Grassmann $f(x)$ and complex $b(x)$ are defined in such a way that all the Fourier amplitudes with $|{\bf p}|>\Lambda$ equal zero identically, i.e. the UV cutoff is setted up only on fields' spatial dependence. The further consideration in this section, except a modifications associated with the presence of a fermion field, will be held in a spirit of the Hugenholtz and Pines paper \cite{Hugenholtz1959} and Ref.~\cite{Hryhorchak2021}. The first step of their approach, which is well-suited for the calculation of thermodynamics of bosons at weak coupling $g_b\to 0$, is the separation of the condensate modes $b(x)=b_0+\tilde{b}(x)$ with the constraint $\int_{L^d}d{\bf r}\tilde{b}=0$. After this, the number of the interaction vertices increases
\begin{eqnarray}\label{}
	& S=\beta L^d\mu_b|b_0|^2+\int dx\,\tilde{b}^*\left\{\partial_{\tau}-\xi_b\right\}\tilde{b}-\beta L^d\frac{g_{b,\Lambda}}{2}|b_0|^4\nonumber\\
	&-\frac{g_{b,\Lambda}}{2}\int dx\,\left\{4|b_0|^2|\tilde{b}|^2+(b^*_0)^2\tilde{b}^2+(\tilde{b}^*)^2b_0^2\right.\nonumber\\
	&\left.+2b_0^*\tilde{b}^*\tilde{b}^2+2(\tilde{b}^*)^2\tilde{b}b_0+|\tilde{b}|^4
	\right\}+\int dx\,f^*\left\{\partial_{\tau}-\xi_f\right\}f\nonumber\\
	& -g_{\Lambda}\int dx\,f^*f\left\{|b_0|^2+b_0^*\tilde{b}+\tilde{b}^*b_0+|\tilde{b}|^2\right\},
\end{eqnarray}
but for the extremely dilute systems the mean field contribution is believed to be the most important. This happens because when $a_b^dn_b, a^dn_f\to 0$ the interaction between atoms is weak and the depletion of the Bose condensate can be neglected. The mean field part of the grand potential in $d>2$ can be written as follows
\begin{eqnarray}\label{Omega_MF}
\Omega_{MF}/L^d=-\mu_b|b_0|^2+\frac{1}{2}g_b|b_0|^4+\Omega^{(0)}_f/L^d+gn_f,
\end{eqnarray}
where $\Omega^{(0)}_f=\sum_{{\bf p}}\xi_f(p)\theta(-\xi_f(p))$ [$\theta(...)$ stands for the Heaviside step function] is the ideal Fermi gas contribution. Importantly that the replacement of the bare couplings $g_{b,\Lambda}$ and $g_{\Lambda}$ by their `observable' counterparts $g_b$ and $g$ can be tracked back by a careful inspection of the diagrams in Figs.~\ref{g_b_fig}, \ref{g_f_fig}. 
Now by means of the minimization of $\Omega_{MF}$ with respect to $b_0$, and using the thermodynamic identities $-\frac{\partial \Omega_{MF}}{\partial \mu_b}=N_b$ and $-\frac{\partial \Omega_{MF}}{\partial \mu_f}=N_f$ (where $N_b$ and $N_f$ is the number of bosons and fermions, respectively), we can relate the leading order chemical potentials at $a_b^dn_b, a^dn_f\ll 1$ to the equilibrium densities $n_b$ and $n_f$
\begin{eqnarray}\label{}
&&\mu_f=\frac{\hbar^2p^2_f}{2m_f}, \ \ p_f=2\sqrt{\pi}\left[\Gamma (d/2+1)n_f\right]^{1/d},\\\label{p_f}
&&\mu_b=g_bn_b+gn_f.
\end{eqnarray}
The stability conditions $\frac{\partial \mu_f}{\partial n_f}>0$, $\frac{\partial \mu_b}{\partial n_b}>0$ and $\frac{\partial \mu_f}{\partial n_f}\frac{\partial \mu_b}{\partial n_b}-\frac{\partial \mu_f}{\partial n_b}\frac{\partial \mu_b}{\partial n_f}>0$ then imply (here we use known fact $\frac{\partial \mu_f}{\partial n_f}=\frac{2}{d}\frac{\mu_f}{n_f}$ for the Fermi energy of an ideal gas)
\begin{eqnarray}\label{MF_stab}
g_b>0, \ \ 
g_b-\frac{d}{2}\frac{g^2n_f}{\mu_f}>0.
\end{eqnarray}
From (\ref{MF_stab}) we learn that the dilute mixtures are always stable, when $g_b>0$ (i.e. repulsive-like), and the absence of the boson-boson interaction necessary leads to the collapse. In the next section, however, we will argue that this is not always the case, and even when $g_b=0$ (but $g>0$), the system possesses a well-defined thermodynamic limit.

\section{Second root}
What was ignored in the analysis of previous section is the fact of potential formation of composite fermions (dimers) consisting exactly of one boson and one fermion. When $g$ is positive and tends to zero (i.e., $a\to 0$ but still $a\gg \Lambda^{-1}$), the dimers become to be more tightly bound and less interacting. For a boson density equal or smaller than the fermion one, the $d=3$ results were previously obtained in \cite{Guidini2014} by means of the $t$-matrix approximation. In order to better understand the phase with the composite particles formed, let us first discuss the balanced case, where the system incorporates an equal (macroscopic) numbers of bosons and fermions.

\subsection{Balanced mixture $n_b=n_f$}
\begin{figure}[h!]
	\centerline{\includegraphics
		[width=0.35
		\textwidth,clip,angle=-0]{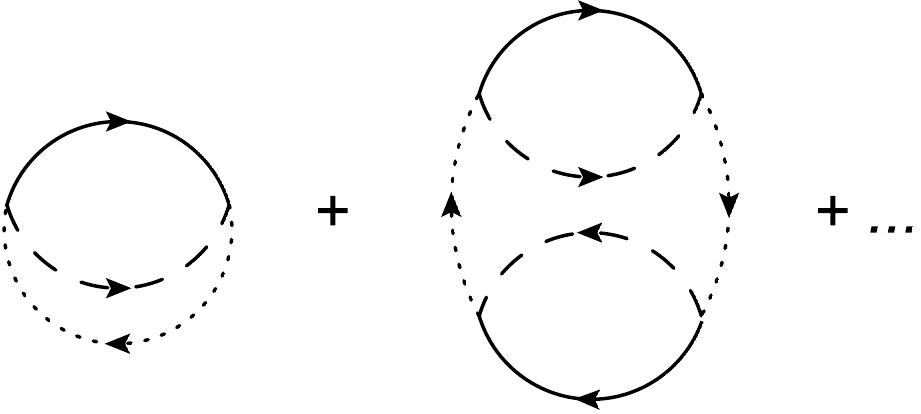}}
	\caption{Diagrammatic representation of the grand potential $\Omega_{dimers}$ for free composite fermions. Doted lines denote the bare propagators of $c$-fermions, $g_{\Lambda}$. Every next term contains additional loop $\Pi_{\Lambda}(P)$ standing between two $g_{\Lambda}$-lines. Up to irrelevant constant term, the series of these circular diagrams sums up in logarithm (\ref{Omega_dimers}).}\label{Omega_dimers_fig}
\end{figure}

A qualitative physical picture of a balanced system with contact interaction can be understood in the Born-Oppenheimer limit by setting the mass of bosons to infinity. The spectrum of a single fermion in a presence of $N_b$ heavy Bose particles possesses exactly $N_b$ bound states with energies close to $\epsilon_B$ \cite{Panochko} in the dilute limit. And in the many-body ground state all these states are occupied by maximum $N_f=N_b$ spinless fermions. In this subsection we neglect for simplicity the inter-bosonic interaction and rewrite the action $S$ introducing composite fermionic (Grassmann) fields $c^*(x)$ and $c(x)$ explicitly
\begin{eqnarray}\label{S_pure_dimers}
S_c=\int dx\,b^*\left\{\partial_{\tau}-\xi_b\right\}b+\int dx\,f^*\left\{\partial_{\tau}-\xi_f\right\}f\nonumber\\
+g^{-1}_{\Lambda}\int dx\,c^*c-\int dx\,\left\{c^*fb+b^*f^*c\right\}.
\end{eqnarray}
The further calculation strategy is predetermined by the physical reasoning. When $a\to 0$ bosons and fermions form a dimers with energy $\epsilon_B$ each, and because of the fermionic nature of these composite particles, their lowest state coincides with the ground state of an almost non-interacting Fermi gas. It is also understood that the phase of a fermionic dimer gas totally excludes both the Bose-Einstein condensation of bosons and free (unbound) fermionic atoms. Indeed, the decay of a single composite fermion into two atoms with a zero momentum each increases the energy of the system by $|\epsilon_B|[1-\mathcal{O}(a^2n^{2/d}_b)]$ in the $an^{1/d}_b\ll 1$ limit. The very same value up to the sign determines the sum of chemical potentials $\mu_b+\mu_f=-|\epsilon_B|[1-\mathcal{O}(a^2n^{2/d}_b)]$. The later observation together with absence of free (non-dimerized) $f$-fermions ($\mu_f\le 0$) immediately decrease the number of non-zero diagrams contributing to the $\Omega$-potential. The simplest series that adequately displays the above-discussed physical picture is given in Fig.~\ref{Omega_dimers_fig}
and results in the thermodynamic potential of free dimers
\begin{eqnarray}\label{Omega_dimers}
\Omega_{dimers}=-\frac{1}{\beta}\sum_{P}\ln t^{-1}(P),
\end{eqnarray}
where the presence of factor $e^{i\nu_p0_+}$ under the sum is assumed and the two-body $t$-matrix is introduced
\begin{eqnarray}\label{}
t^{-1}(P)=g^{-1}_{\Lambda}+\Pi_{\Lambda}(P),
\end{eqnarray}
with the boson-fermion loop given by $\Pi_{\Lambda}(P)=\frac{1}{\beta L^d}\sum_K G_b(K)G_f(P-K)$. Performing the Matsubara frequency integration and compensating the UV divergences with the help of Eq.~(\ref{g_def}), we obtain a finite result
\begin{eqnarray}\label{t_P}
& t^{-1}(P)=g^{-1}\left\{1-\frac{\left[\frac{\hbar^2p^2}{2M}-\mu_f-\mu_b-i\nu_p
\right]^{d/2-1}}{|\epsilon_B|^{d/2-1}}\right\},
\end{eqnarray}
here $M=m_f+m_b$ denotes the mass of a composite fermion. Then, the calculation of integrals in (\ref{Omega_dimers}) and use of thermodynamic identities leave us with an expression relating the density of dimers to the chemical potentials of an extremely dilute Bose-Fermi system $\mu_b+\mu_f=\epsilon_B+\frac{\hbar^2p^2_f}{2M}$, with the Fermi wave-vector $p_f$ defined in (\ref{p_f}). 

\begin{figure}[h!]
	\centerline{\includegraphics
		[width=0.45
		\textwidth,clip,angle=-0]{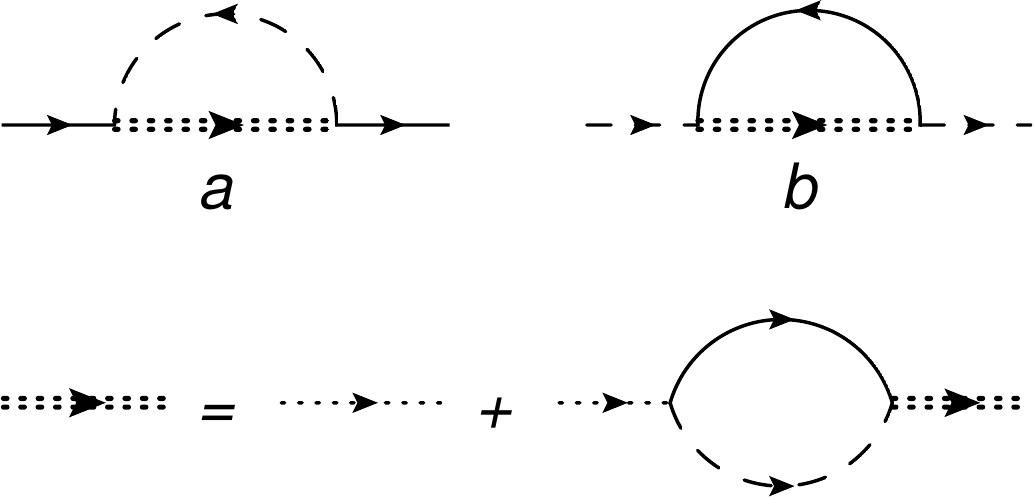}}
	\caption{Leading-order self-energies $\Sigma_f(K)$ (diagram a) and $\Sigma_b(K)$ (diagram b) of the fermionic and bosonic propagators, respectively. Double dotted line stands for a dressed (in simplest way) dimer propagator $t(P)$ [see Eq.~(\ref{t_P})], which possesses a pole in the upper complex half-plane at the finite densities of composite particles.}\label{Sigma_f_b_fig}
\end{figure}

Same results can be equivalently obtained by inclusion the simplest self-energy corrections (see Fig.~\ref{Sigma_f_b_fig})
to the Green's functions of bosons
\begin{eqnarray}\label{Sigma_b}
&&\Sigma_{b}(K)=\nonumber\\
&&-\frac{1}{ L^d}\sum_{|{\bf p}|\le p_f}\frac{\frac{g}{d/2-1}|\epsilon_B|}{\xi_{f}(|{\bf p}-{\bf k}|)+\frac{\hbar^2(p^2_f-p^2)}{2M}+i\omega_k},
\end{eqnarray}
and fermions $\Sigma_{f}(P)=-\Sigma_{b}(P)|_{m_b\leftrightarrow m_f, \mu_b\leftrightarrow \mu_f}$. Besides thermodynamic properties these expressions allow the calculations of the particle momentum distributions. Particularly, for bosons we obtain
\begin{eqnarray}\label{N_b}
&&N_{b}(k)=-\frac{1}{\beta}\sum_{\omega_k}[G_b(K)]^2\Sigma_{b}(K)=\nonumber\\
&&\frac{1}{L^d}\sum_{|{\bf p}|\le p_f}\frac{\frac{g}{d/2-1}|\epsilon_B|}{\left[\xi_{f}(|{\bf p}-{\bf k}|)+\xi_b(k)+\frac{\hbar^2(p^2_f-p^2)}{2M}\right]^2},
\end{eqnarray}
which is well approximated by the squared Lorentzian profile in the region of applicability. A qualitatively similar behavior is intrinsic for the momentum distribution of fermions $N_f(p)$. In fact, $N_{b}(k)$ and $N_f(p)$ coincide up to the leading order in $a^dn_{b,f}$. It is straightforward to show that $\frac{1}{L^d}\sum_{{\bf k}}N_{b}(k)=n_b$ and that the high-momentum tale is consistent with the universal Tan's behavior \cite{Tan} $N_{b}(k\to \infty)\sim  \mathcal{C}_b/k^4$, where the contact parameters $\mathcal{C}_{b,f}=\frac{(4\pi)^{d/2}}{\Gamma(2-d/2)}\frac{n_{b,f}}{a^{4-d}}$ are equal to each other in the limit $a^dn_{b,f}\ll 1$.

Formulas (\ref{Sigma_b}) and (\ref{N_b}) (and their fermionic equivalents) tell us about the stability of the system in balanced phase. Two inequalities $\frac{\partial n_f}{\partial \mu_f}>0$ and $\frac{\partial n_b}{\partial \mu_b}>0$ are always satisfied, but the Jacobian $\frac{\partial(n_b, n_f)}{\partial(\mu_b, \mu_f)}$ equals zero identically signalling the threshold of the thermodynamic instability. This is not actually crucial for the fate of the system, because Eq.~(\ref{Omega_dimers}) totally neglects the interaction between the composite particles, which can be either repulsive or attractive, but importantly the odd-wave type. A naive scaling of coupling constant in the $p$-wave channel reads $a^{7-d}$ (i.e. this interaction is negligibly small in the adopted approximation). The repulsion between dimers necessary stabilizes the system, while the ground state for the attractive interaction is the superfluid. The transition temperature of the superfluidity emergence in the system is exponentially suppressed and that is why the above analysis is valid even at very low temperatures.

\subsection{$n_f>n_b$}\label{nf_nb}
In previous subsection we calculated $\mu_b+\mu_f$ for fully balanced mixture without specifying the chemical potentials of boson and fermions separately, and naively one may think that $\mu_f=\mu_b$. But this is not true. In order to understand this let us put in the system of dimers one more fermion. Because there are no free bosons, this atom should occupy the energy level in the continuum. The same happens for the second, third and other fermions that will be thrown in the system. Now if we put in a slightly imbalanced mixture ($n_f> n_b$) a single boson, it will bound a fermion from the continuum, which in turn, will decrease the energy of the system by $\approx |\epsilon_B|$. This finally leads us to the conclusion that $\mu_f=0$ (up to the finite-size corrections) and $\mu_b=\epsilon_B+\frac{\hbar^2p^2_f}{2M}$ for a balanced mixture in the dilute limit.

\begin{figure}[h!]
	\centerline{\includegraphics
		[width=0.475
		\textwidth,clip,angle=-0]{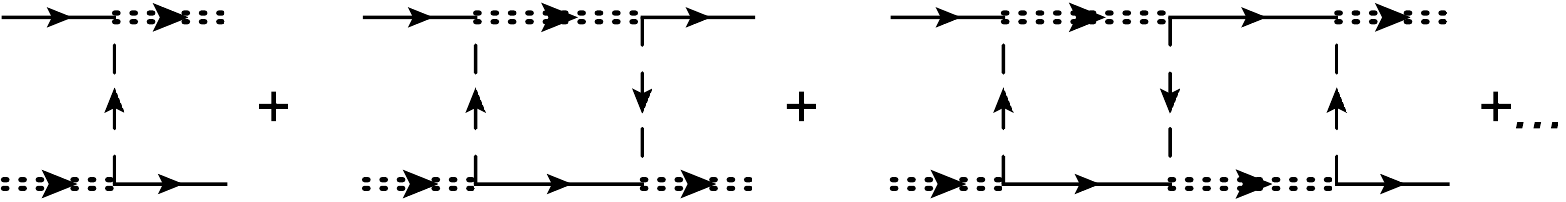}}
	\caption{Fermion-dimer scatterings. The series is exact in vacuum and can be summed up by means of linear integral equation for the fermion-dimer vertex $\mathcal{T}_{fc}(P_1;P_2|P_2';P_1')$ in a $d+1$ momentum space.}\label{fermion_dimer_fig}
\end{figure}

The imbalanced mixture in the region of validity of our discussion consists of $N_b$ composite particles and $N_f-N_b$ free fermionic atoms. Performing the path integral, we obtain (recall that now $\mu_f>0$)
\begin{eqnarray}\label{Omega_dim_ferm}
\Omega_{fc}=\Omega^{(0)}_f-\frac{1}{\beta}\sum_{P}\ln [t^{-1}(P)+\Delta t^{-1}(P)],
\end{eqnarray}
the grand potential of the Fermi-Fermi mixture. The finite density of the non-dimerized fermions is responsible for the correction to the boson-fermion scattering loop
\begin{eqnarray}\label{Delta_t}
\Delta t^{-1}(P)=\frac{-1}{L^d}\sum_{|{\bf p}'|\le p_f}
\frac{1}{\xi_b(|{\bf p}'-{\bf p}|)+\xi_f(p')-i\nu_p}.
\end{eqnarray}
First of all, this correction is small in comparison to $t^{-1}(P)$, and can be neglected at initial stage of calculations. The resulting ground state energy includes two terms each referred to the ideal Fermi gas. The first one contains particles of the mass  $m_f$ and density $n_f-n_b$, while the second one $M$ and $n_b$. Secondly, in order to find out the nature of its impact it is enough to expand the logarithm in the second term of $\Omega_{fc}$. Then, simple computations of integrals in the leading order in parameter $a^dn_{b,f}\ll 1$, yield the correction to the energy density $\frac{g}{d/2-1}n_b(n_f-n_b)$ of the system. Thus, the presence of $\Delta t^{-1}(P)$ describes the simplest two-body interaction between composite particles and free fermions. It should be noted that $\frac{1}{\beta}\sum_{\nu_p}e^{i\nu_p0_+}t(P)$ differs from the momentum distribution of the non-interacting composite particles by a constant prefactor $\frac{g|\epsilon_B|}{d/2-1}$. Furthermore, because this residue of $t(P)$ is large, by a careful inspection of diagrams one readily concludes that linear in $g$ corrections to the energy density are produced by the whole series of the fermion-composite-particle scattering processes presented in Fig.~\ref{fermion_dimer_fig} (where the first diagram is responsible for the above-mentioned linear in $\Delta t^{-1}(P)$ correction).
Then, in order to calculate the linear-$g$ corrections to thermodynamics of a considered system one has to connect free lines in every diagram. In the limit of extreme diluteness of the system, we can approximately rewrite the energy density correction as follows
\begin{eqnarray}\label{}
\frac{1}{(\beta L^{d})^2}\sum_{P, P'}G_f(P)t(P')\mathcal{T}_{fc}(P;P'|P';P)\nonumber\\
\approx g_{fc}n_b(n_f-n_b),
\end{eqnarray}
where the effective coupling constant $g_{fc}=\mathcal{T}_{fc}(P;P'|P';P)|_{i\nu_p\to \xi_f(p), i\nu_{p'}\to \frac{\hbar^2(p'^2-p_f^2)}{2M}}$ (and then ${\bf p}, {\bf p}'\to 0$ and neglect terms of order $p_fa$). The above simplifications suggest that we only have to compute the on-shell fermion-dimer scattering amplitude at zero momentum in vacuum
\begin{eqnarray}\label{}
	&& g_{fc}=\frac{g|\epsilon_B|}{d/2-1}\mathcal{T}_{fc}(p\to 0),\\ 
	&& \mathcal{T}_{fc}(p)= \mathcal{T}_{fc}(P;-P|0;0)|_{i\nu_p\to -\xi_f(p)},
\end{eqnarray}
to reveal the dependence of the effective coupling $g_{fc}$ on the spatial dimensionality and fermion-boson mass ratio. 

\begin{figure}[h!]
	\centerline{\includegraphics
		[width=0.475
		\textwidth,clip,angle=-0]{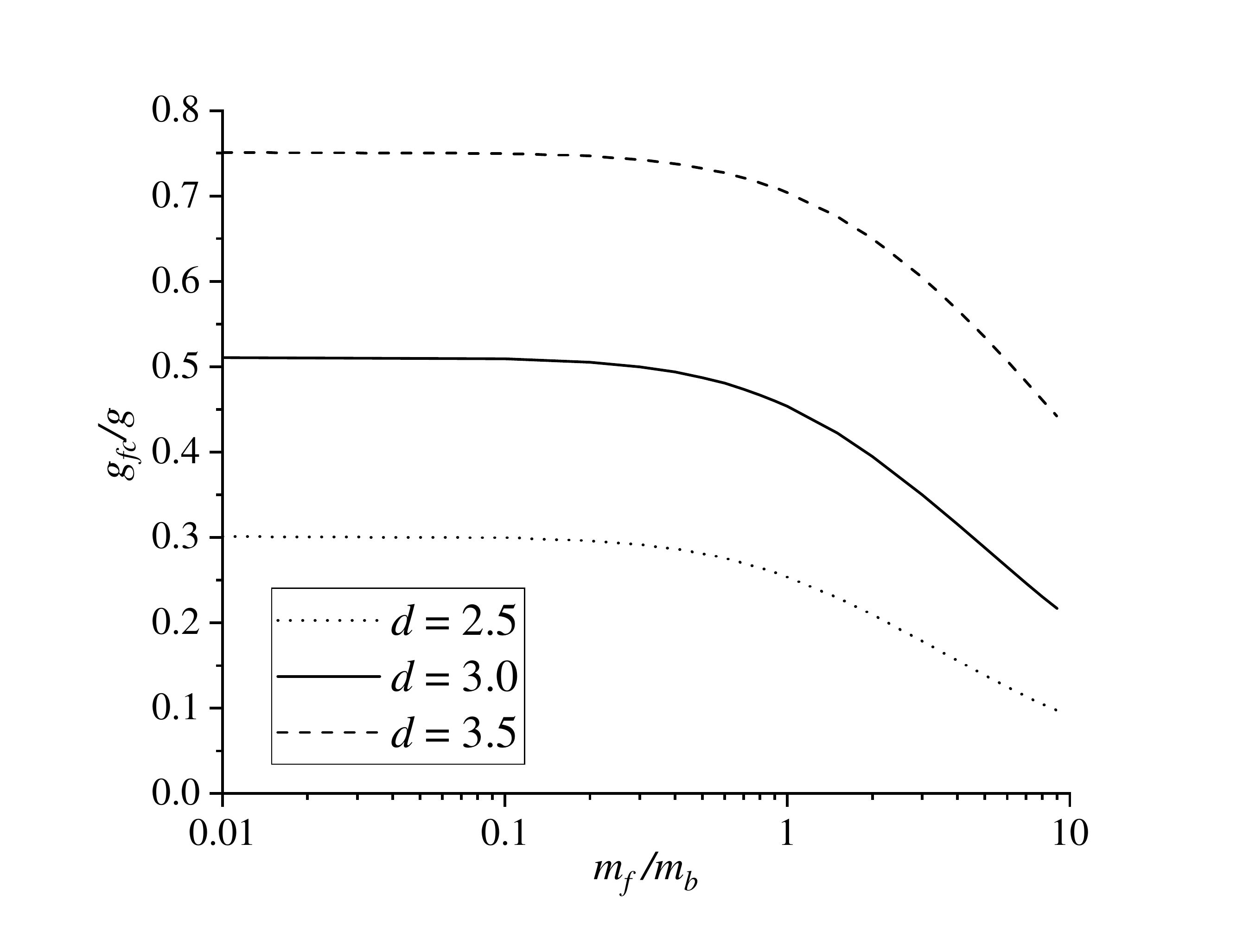}}
	\caption{Fermion-dimer effective coupling constant (in unit of $g$) for various mass ratios.}\label{g_fc_fig}
\end{figure}

The dimensionless function $X(p)=|\epsilon_B|\mathcal{T}_{fc}(p/a)$ satisfies the non-homogeneous linear integral equation
\begin{eqnarray}\label{X_p}
X(p)=\frac{1}{1+p^2}+\frac{2\sin\left(\pi d/2\right)}{\pi}\int \frac{dss^{d-1}}{1+p^2+s^2}\nonumber\\
\times\frac{_2F_1\left(1/2,1; d/2;\left(\frac{2m_fps/M}{1+p^2+s^2}\right)^2\right)}
{\left(1+\frac{m_r}{M_f}s^2\right)^{d/2-1}-1}X(s),
\end{eqnarray}
with $_2F_1(a,b;c;z)$ being hypergeometric function \cite{Abramowitz} and $M_f=\frac{Mm_f}{M+m_f}$ the fermion-dimer reduced mass. We solved this equation numerically for the infinite UV cutoff and arbitrary sets of $d$ and $m_f/m_b$, and then plotted in Fig.~\ref{g_fc_fig} 
the induced fermion-dimer coupling $g_{fc}$. The latter is found to be the same order magnitude for various mass ratios, but always positive definite. Thus, the imbalanced $n_f>n_b$ dilute Bose-Fermi mixture with the composite fermions formed satisfies all the stability conditions.

\subsection{$n_b>n_f$}
Let us put in the system more Bose atoms than fermions. Suppose then that there are exactly $N_f$ composite dimers and all $N_b-N_f$ non-bound bosons fell out into the zero-momentum Bose condensate. Because now the immersion of additional fermion decreases the energy of the system by $\approx |\epsilon_B|$, we readily conclude that chemical potential $\mu_f$ in this phase (if it exists, of course) is negative and of order of a two-body binding energy for dilute systems. Immersion of one more boson, instead, does not change the energy of dilute system sufficiently. However, the calculations of the thermodynamic potential by utilizing action (\ref{S_pure_dimers}) with the separated condensate mode of fields $b(x)$
\begin{eqnarray}\label{Omega_dim_bos}
\Omega_{bc}=-L^d\mu_b|b_0|^2-\frac{1}{\beta }\sum_{P}\ln \left[t^{-1}(P)-|b_0|^2G_f(P)\right],
\end{eqnarray}
yield for the bosonic chemical potential $\mu_b=\frac{gn_f}{d/2-1}$ (it is calculated by extremizing $\Omega_{bc}$ with respect to $b_0$). This particularly means that a dilute mixture consisting of the majority of bosons and minority of fermions is unstable toward collapse in absence of the direct inter-boson repulsion. The issue can be cured by restoring in action (\ref{S_pure_dimers}) the two-body interaction between bosons with a small positive $g_b$. In addition, we have to assume that the non-dimerized bosons are in the so-called `upper branch' phase described in Sec.~\ref{weak-coupling_sec} and do not form the two-body bound states. The occurrence of such a bosonic dimers necessarily leads to instability provided by the attractive nature of their interaction. Having discussed all physical conditions for the creation of the Bose-Fermi mixture with the boson-fermion dimers formed in a case $n_b>n_f$, and assuming $\mu_b$ being small in comparison to $|\epsilon_B|$ (which will be confirmed {\it a posteriori}) we add to $\Omega$-potential (\ref{Omega_dim_bos}) term $L^d\frac{g_{b}}{2}|b_0|^4$, which is believed to stabilize the system. Note that in following, this mean-field inter-boson interaction is the only preserved. It is a well-justified approximation when $g_b\ll g$.

\begin{figure}[h!]
	\centerline{\includegraphics
		[width=0.475
		\textwidth,clip,angle=-0]{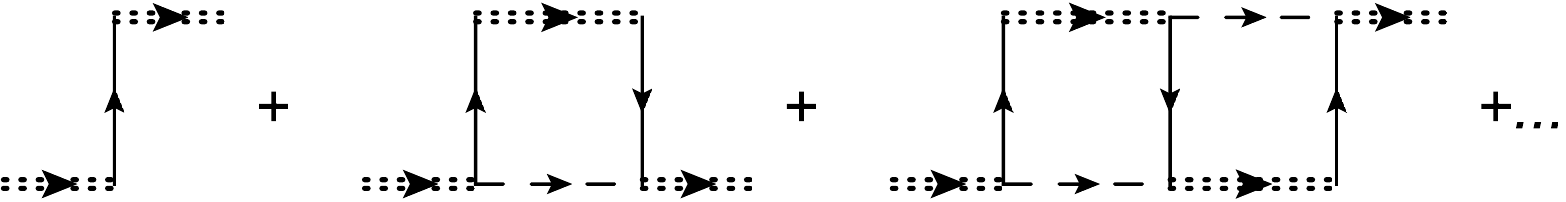}}
	\caption{Series of diagrams describing the condensed-boson-dimer scatterings (the condensate lines are not shown). Note that by encircling the first diagram we get a term proportional to $|b_0|^2$ in the logarithm of thermodynamic potential (\ref{Omega_dim_bos}). The summation of the series can be done as follows: starting from the second one we can cut first two horizontal lines in every diagram obtaining the series for the three-legged vertex with two incoming (bosonic and dimer) lines and one outgoing (dimer) line. This vertex with the on-shell conditions applied is determined by the closed form linear integral equation, which greatly simplifies (\ref{Tau_bc}), when the momentum of the dimer vanishes.}\label{boson_dimer_fig}
\end{figure}

A more careful inspection of the second term in (\ref{Omega_dim_bos}) provides the possibility in a dilute limit we can expand the logarithm leaving only linear in $|b_0|^2t(P)G_f(P)$ contribution. Moreover, likewise the situation from the previous subsection, the same type (and importantly, same order) corrections to thermodynamics are generated by the infinite series of diagrams depicted in Fig.~\ref{boson_dimer_fig}, 
and which contain two condensate lines and can be treated as the self-energy corrections to the propagator of composite particles. Technically this issue is very similar to the one from subsection~\ref{nf_nb}: the series in the dilute limit for the on-shell vertex is summed up by the non-homogeneous integral equation
\begin{eqnarray}\label{Tau_bc}
&&	\mathcal{T}_{bc}(k)=-\frac{1}{\xi_f(k)+\xi_b(k)}\nonumber\\
&&	-\frac{1}{L^d}\sum_{|{\bf p}|<\Lambda}\frac{t(P)|_{i\nu_p \to -\xi_b(p)}\mathcal{T}_{bc}(p)}{\xi_f(|{\bf k}+{\bf p}|)+\xi_b(p)+\xi_b(k)},
\end{eqnarray}
the leading-order correction to energy density
\begin{eqnarray}\label{}
g_{bc}|b_0|^2n_f,
\end{eqnarray} 
requires only the zero-momentum limit of a solution $g_{bc}=\frac{g|\epsilon_B|}{d/2-1}\mathcal{T}_{bc}(k\to 0)$.
However, the solution itself is much more complicated than that of Eq.~(\ref{X_p}) due to attractive nature of the boson-dimer interaction. It also possesses the explicit $\Lambda$-dependence. Even more, in this limit the effective field theory described by the action (\ref{S_pure_dimers}) is incomplete to capture \cite{BHvK_99_1,BHvK_99_2} the three-body physics appropriately and requires the introduction of a bare boson-dimer interaction 
\begin{eqnarray}\label{}
\Delta S=-g_{bc,\Lambda}\int dx\,|b|^2c^*c,
\end{eqnarray}
where the naive dependence of the bare coupling $g_{bc,\Lambda}=\frac{2m_r}{\hbar^2\Lambda^2}\hat{g}_{bc,\Lambda}$ (here $\hat{g}_{bc,\Lambda}$ is dimensionless) on UV cutoff is found by the dimensional analysis. Having supplemented the action by an additional term, we also have to modify the integral equation (\ref{Tau_bc}), where the replacements $\frac{1}{\xi_f(k)+\xi_b(k)}\to \frac{1}{\xi_f(k)+\xi_b(k)}-g_{bc,\Lambda}$ in the non-homogeneous term and $\frac{1}{\xi_f(|{\bf k}+{\bf p}|)+\xi_b(p)+\xi_b(k)}\to \frac{1}{\xi_f(|{\bf k}+{\bf p}|)+\xi_b(p)+\xi_b(k)}-g_{bc,\Lambda}$ under the sum, should be made. Then, in order to calculate the running boson-dimer coupling $\hat{g}_{bc,\Lambda}$, we demand that the solution $\mathcal{T}_{bc}(k\to 0)$ of (\ref{Tau_bc}) is independent on the UV cutoff. This renormalization-group scheme can be
realized only numerically \cite{Nakayama}, but there is a simple and quite accurate analytic approximation \cite{Mohapatra}. 

\begin{figure}[h!]
	\centerline{\includegraphics
		[width=0.475
		\textwidth,clip,angle=-0]{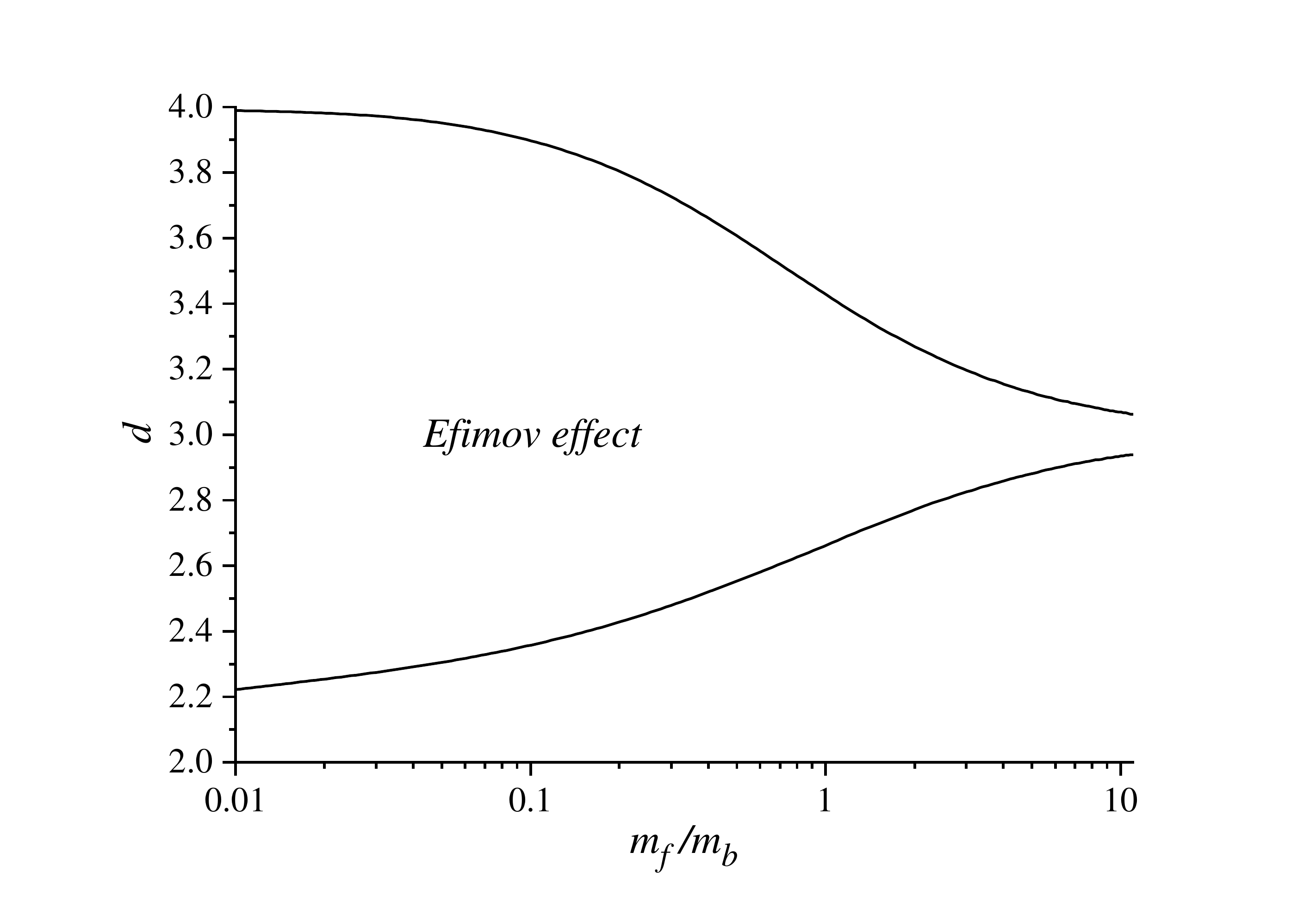}}
	\caption{`Window' for the Efimov effect in the system with two bosons and one fermion. Because we assume no boson-boson interaction, the induced effective attraction between them exhausts when the mass of a fermionic atom is large.}\label{Efimov_window_fig}
\end{figure}

The attractive character of the boson-dimer interaction provides the formation of the three-body (two bosons + one fermion) bound states. The emergence of the universal Efimov trimers \cite{Efimov,Naidon} crucially depends on the spacial dimensionality \cite{Nielsen_et_al} and mass ratios of particles \cite{Rosa_et_al}. In our case, when two bosons are almost non-interacting but interact with a fermionic atom, the `window' (see Fig.~\ref{Efimov_window_fig})
for the Efimov physics is given by a pair of solutions (with $\eta=0$)
\begin{eqnarray}\label{}
& 1+\frac{\sin\left(\frac{\pi d}{2}\right)}{\cos\left(\frac{\pi \eta}{2}\right)}\, _2F_1\left(\frac{d-1-\eta}{2},\frac{d-1+\eta}{2}; \frac{d}{2};\left(\frac{m_b}{M}\right)^2\right)=0.
\end{eqnarray}
In fact, at fixed $m_f/m_b$ and $d$, the above equation determines the exponent $\eta$ of the power-law behavior of $\mathcal{T}_{bc}(k)$ in the so-called scaling region $1/a\ll k\ll \Lambda$. Fully imaginary $\eta=i\eta_0$ solutions correspond to the Efimov effect. The lowest three-body bound-state energy level scales like $|\epsilon_3|\propto\frac{\hbar^2\Lambda^2}{M+m_b}$, and positions of all highest levels $\epsilon^{(n)}_3$ in the interval $[\epsilon_3, \epsilon_B]$ can be found by the discrete scale invariance $\epsilon^{(n+1)}_3/\epsilon^{(n)}_3=e^{-2\pi/\eta_0}$ ($n\gg 1$). Actually, the occurrence of such a deep trimers is demolishing for the dimers. Indeed, the trimers also form a thermodynamically stable almost ideal Fermi gas in the considered dilute limit, but a huge binding energy makes them more energetically preferable. An interesting phase with the coexistence of dimers and trimers in the system can only occur when $2\epsilon_B\approx \epsilon^{(n)}_3$, but this case is out the scope of present study and deserves a separate publication.

\begin{figure}[h!]
	\centerline{\includegraphics
		[width=0.475
		\textwidth,clip,angle=-0]{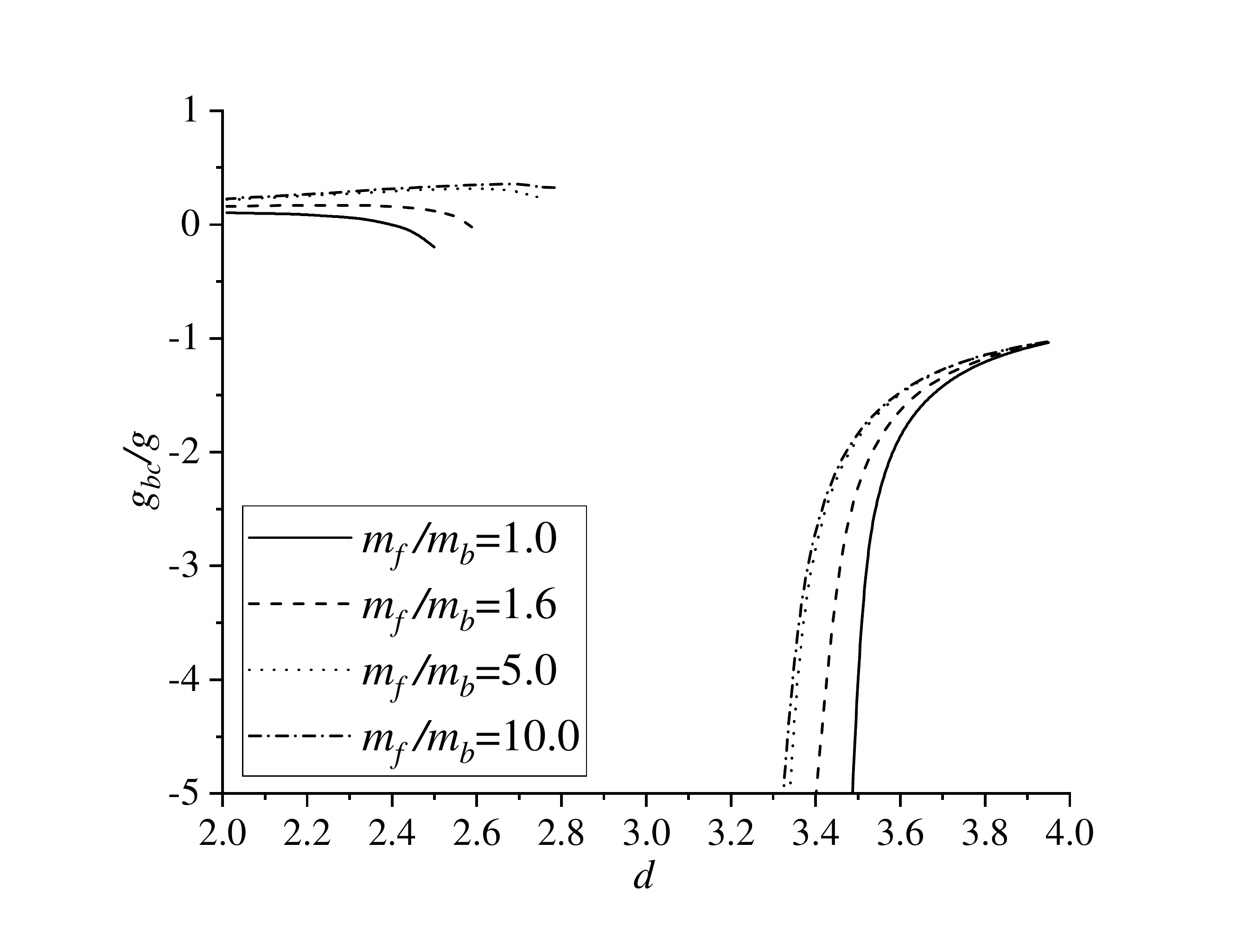}}
	\caption{Boson-dimer effective coupling constant (in unit of $g$) in various dimensions outside the Efimov `window' (Fig.~\ref{Efimov_window_fig}).}\label{g_bc_fig}
\end{figure}

Summarizing the above discussion, we have found out that the considered phase with $N_f$ dimers and $N_b-N_f$ condensed bosons interacting through the short-ranged potential is at least metastable because of the enormously large three-body binding energy, while the true ground state supports a maximal number of trimers. For same reasons the state described in Sec.~\ref{weak-coupling_sec} also is not the one that minimizes the energy of the Bose-Fermi mixture. The lifetime of the system initially prepared with the dimers formed, however, has to be comparatively large because of the large energy difference between these two states. There is a particular interest in obtaining the stability condition of the dimer state outside the Efimov `window'. The main ingredient of these calculations is the boson-dimer coupling $g_{bc}=\frac{g|\epsilon_B|}{d/2-1}\mathcal{T}_{bc}(0)$, which we calculated (see Fig.~\ref{g_bc_fig})
by solving Eq.~(\ref{Tau_bc}) (with $g_{bc,\Lambda}$ inserted) numerically outside the Efimov `window' in the limit $\Lambda\to \infty$. This procedure automatically provides the computations of the UV fixed point for the running coupling $\hat{g}_{bc,\Lambda}$. With $g_{bc}$ in hand, we can obtain the grand potential and all other thermodynamic functions in the adopted approximation  $a^dn_{b,f}\to 0$. Then, from the thermodynamic identities $-\frac{\partial \Omega/L^d}{\partial \mu_{b,f}}=n_{b,f}$, we get
\begin{eqnarray}\label{}
n_b=|b_0|^2+n_f.
\end{eqnarray}
Minimizing $\Omega$ with respect to a condensate density and taking into account the abode equality, we obtain
\begin{eqnarray}\label{}
\mu_b=g_b(n_b-n_f)+g_{bc}n_f.
\end{eqnarray}
These two equations together fix the chemical potential of fermions
\begin{eqnarray}\label{}
\mu_f+|\epsilon_B|=\frac{\hbar^2p_f^2}{2M}-g_b(n_b-n_f)-g_{bc}(2n_f-n_b).
\end{eqnarray}
The thermodynamic stability of the system with dimers and fully condensed unbound bosons against collapse implies
\begin{eqnarray}\label{bc_stab}
g_b>0, \ \ 
g_b-\frac{d}{2}\frac{g_{bc}^2n_f}{\hbar^2p_f^2/2M}>0,
\end{eqnarray}
which looks like a standard weak-coupling mean-field condition (\ref{MF_stab}), except for the effective boson-dimer coupling $g_{bc}$ and chemical potential $\hbar^2p_f^2/2M$ of an ideal dimer gas.

\section{Summary}
In conclusion, we have studied, by means of the effective field theory approach, the dimer phase of a dilute two-component Bose-Fermi mixture in a spatial dimensions between $d=2$ and $d=4$. Utilizing two simple principles, namely, the energy minimization and condition for the thermodynamic stability of a mixture against either collapse or phase separation into the pure Bose and Fermi gases, we have argued that there are some compositions of a system that leave it stable even when the inter-boson interaction vanishes. This obviously contradicts the standard weak-coupling phase of the Bose-Fermi mixture with almost undepleted Bose condensate and weakly-interacting spin-polarized fermions. Considering the fermion-dimer and the boson-dimer scattering properties, we have calculated the effective coupling constants, which are responsible for the leading-order shifts of the thermodynamic functions of the dilute system. Neglecting the boson-boson interaction, we have found out the conditions for the emergence of Efimov effect in a system of two bosonic and one fermionic atoms. The existence of such a tightly bound trimers makes the dimer matter metastable, but a large energy difference between phases with either dimers or trimers formed inspires hope for the observation of dimers in the dilute Bose-Fermi mixtures in experiments and numerical simulations.

\begin{center}
	{\bf Acknowledgements}
\end{center}
We thank Dr.~I. Pastukhova for invaluable comments on the manuscript.
This work was partly supported by Project No.~0122U001514 from the Ministry of Education and Science of Ukraine.

\end{document}